\documentclass[aps,prl,twocolumn,showpacs]{revtex4}
\usepackage{graphicx,latexsym}
\pdfoutput=1

\begin{document}
\title{Determination of the Equation of State of a Polarized Fermi Gas at Unitarity}

\author{Yong-il Shin}\email{yishin@mit.edu}

\affiliation{Department of Physics, MIT-Harvard Center for Ultracold
Atoms, and Research Laboratory of Electronics, Massachusetts Institute of
Technology, Cambridge, Massachusetts, 02139}

\date{\today}

\begin{abstract}
We report on the measurement of the equation of state of a two-component
Fermi gas of $^6$Li atoms with resonant interactions. By analyzing the
\textit{in situ} density distributions of a population-imbalanced Fermi
mixture reported in the recent experiment [Y. Shin \textit{et al.},
Nature \textbf{451}, 689 (2008)], we determine the energy density of a
resonantly interacting Fermi gas as a function of the densities of the
two components. We present a method to determine the equation of state
directly from the shape of the trapped cloud, where the fully-polarized,
non-interacting ideal Fermi gas in the outer region provides the absolute
calibration of particle density. From the density profiles obtained at
the lowest temperature, we estimate the zero-temperature equation of
state.
\end{abstract}

\pacs{03.75.Ss, 03.75.Hh, 51.30.+i}

\maketitle

Interacting fermions are a paradigm of modern physics, encompassing
superconductivity and superfluidity. One interesting situation arises
when the constituents interact resonantly, i.e., the scattering length
for the free fermions diverges. At this so-called unitarity limit, the
behavior of the system becomes universal, being independent of the nature
of the interactions. Ultracold atomic Fermi gases near Feshbach
collisional resonance present a model system for studying strongly
interacting fermions~\cite{GPS07}. Recently, the phase diagram of a
two-component Fermi gas with resonant interactions has been experimentally
established~\cite{SSS08}, showing that at zero temperature the system
undergoes a first-order quantum phase transition from a fully-paired
superfluid to a partially-polarized normal gas when the imbalance between
the two spin components exceeds a critical value, called the
Chandrasekhar-Clogston limit of superfluidity~\cite{Chandra62,CLO62}.

The nature of the partially-polarized normal phase, however, is still a
subject of investigation. The spectral shift observed in the minority rf
excitation spectrum has been interpreted as the existence of `pairing' in
the normal phase~\cite{SSS07}, but several theoretical studies for a
highly polarized system, e.g. a single minority atom in a majority Fermi
sea, suggest that the system is well-described as a normal Fermi liquid,
where the minority atoms are associated with weakly interacting
quasiparticles~\cite{LRG06, CRL07, PS07}. This picture seems to be
supported by the experimental observation that the shape of the minority
cloud in the normal phase is similar to a free Fermi
gas~\cite{ZSS06b,SSS08}. It has been speculated that exotic pairing
states might exist in the partially-polarized phase~\cite{BFS06}.

In this Letter, we determine the equation of state of a polarized Fermi
gas at unitarity from the the \textit{in situ} density profiles of a
population-imbalanced Fermi mixture confined in a harmonic trap. Since
the variation of the external trapping potential across the sample scans
the chemical potential, in principle, the density information of a single
sample contains the whole equation of state~\cite{Chevy06, BF07}. We
present a method to determine the equation of state directly from the
shape of the trapped cloud. Because of its exactly-known thermal
properties, a fully-polarized, non-interacting ideal Fermi gas in the
outer region provides the absolute density calibration. The equation of
state of a polarized Fermi gas can be parameterized using a normal Fermi
liquid description, which includes the binding energy of a single
minority atom resonantly interacting with a majority Fermi sea, the
effective mass of the quasiparticles, and its correction. This work is
the first quantitative study of the thermodynamic properties of the
polarized normal state with strongly interactions, finding reasonable
agreement with recent calculations~\cite{Chevy06,LRG06,CRL07,BF07,PS07}.

For infinite scattering length, the unitarity limit implies that all
interaction energies scale with the Fermi energies of the components
$\varepsilon_{Fi}=\hbar^2/2m(6\pi^2n_i)^{2/3}$~\cite{Ho04}, where $\hbar$
is the Planck constant divided by $2\pi$, $m$ is the particle mass, $n_i$
is the density of component $i$, and $i=1,2$. As a result, a simple
dimensional scaling argument implies that the energy density
$\mathcal{E}(n_1,n_2)$ of a two-component Fermi gas can be parameterized
as
\begin{equation}
   \mathcal{E}(n_1,n_2)=\frac{3}{5} \alpha [n_1 g(x)]^{5/3},
\end{equation}
introducing a dimensionless universal function $g(x)$~\cite{BF07}, where
$\alpha=(6\pi^2)^{2/3}\hbar^2/2m$ and $x=n_2/n_1$ is the density ratio of
the two components. Without loss of generality, $0 \leq x \leq 1$ due to
the symmetry of the two components. From the chemical potential relation
$\mu_i=\partial \mathcal{E} /\partial n_i$, the universal function can be
expressed as
\begin{equation}
 g(x)^{5/3}= \frac{\mu_1 + x \mu_2}{\alpha n^{2/3}_1}=
\frac{\mu_1}{\varepsilon_{F1}} (1+ xy),\label{e:gx}
\end{equation}
where $y=\mu_2/\mu_1$ is the chemical potential ratio of the two
components.

The main result of this paper is the measurement of the universal
function $g(x)$ for a resonantly interacting Fermi gas. When a Fermi
mixture is confined in a harmonic trap, $V(r)\propto r^2$, the local
chemical potential is given as $\mu_i(r)=\mu_{i0}-V(r)$, where $\mu_{i0}$
is the global chemical potential with respect to the bottom of the
trapping potential. The global chemical potential of the majority
component $\mu_{10}$ is determined from the radius of the majority cloud
$R_1$, i.e., $\mu_{10}=V(R_1)$. Then, the chemical potential ratio
$y(r)=\mu_2(r)/\mu_1(r)$ is given as
\begin{equation}
 y(r)=\frac{y_0-r^2/R^2_1}{1-r^2/R^2_1},
\end{equation}
where $y_0 = \mu_{20}/\mu_{10}$. With population imbalance, i.e. $y_0<1$,
$x(r)$ and $y(r)$ vary over the sample. From Eq.~(\ref{e:gx}), the
spatial correlation of the local particle densities and the local
chemical potentials in a trapped sample determines $g(x)$ under the local
density approximation.

\begin{figure}
\begin{center}
\includegraphics[width=2.3in]{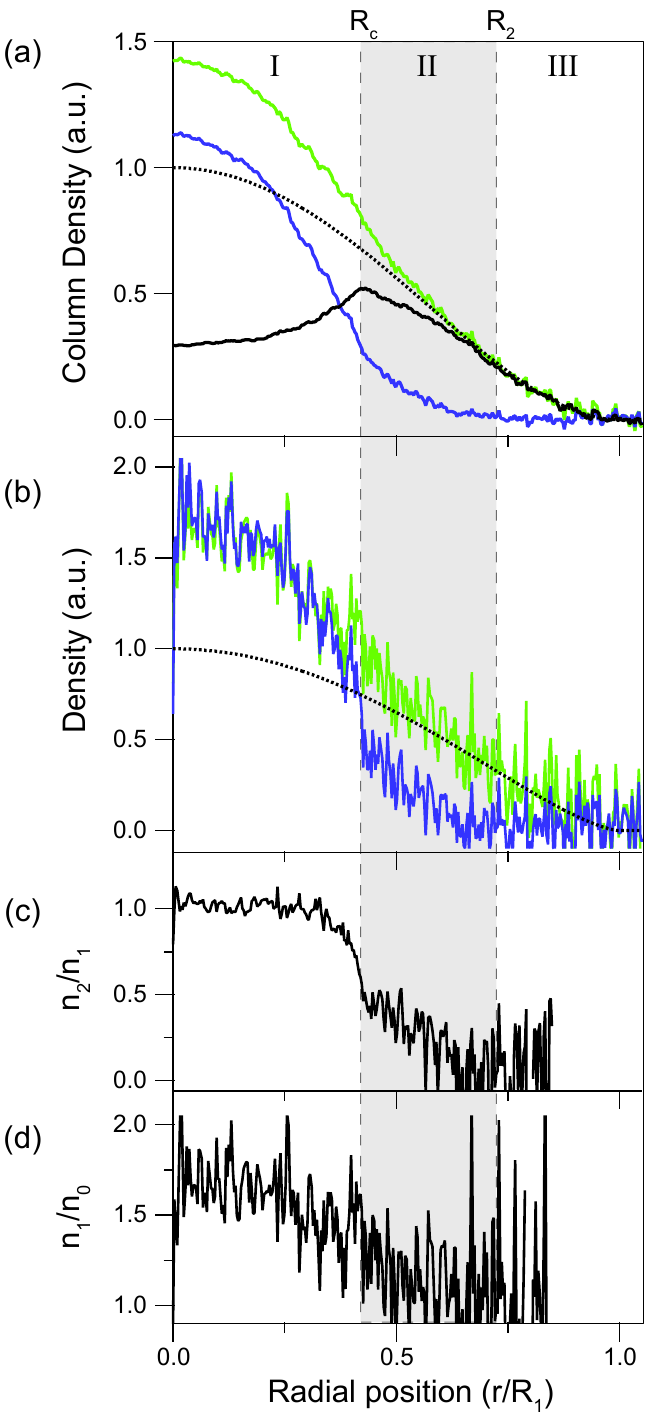}
\caption{(Color online) Spatial profiles of a population-imbalanced Fermi
mixture confined in a harmonic trap at unitarity. (a) The averaged column
density profiles and (b) the reconstructed 3D profiles at the lowest
temperature from Ref.~\cite{SSS08} (green (gray): majority, blue (dark
gray): minority, black: difference). $R_1$ and $R_2$ are the radii of the
majority (spin 1) and the minority (spin 2) cloud, respectively. The core
radius $R_c$ was determined as the kink and peak position in the column
density difference. The black dotted line in (a) is a zero-temperature
Thomas-Fermi (TF) distribution fit to the outer region ($r>R_2$) of the
majority column density profile and the black dotted line in (b) is the
corresponding 3D distribution, $n_0$. The sample has three regions: (I)
The superfluid core region ($0\leq r <R_c$), (II) the partially-polarized,
intermediate region ($R_c< r <R_2$), and (III) a fully-polarized
non-interacting outer region ($R_2< r <R_1$). (c) The density ratio of the
two components, $x(r)=n_2/n_1$. (d) The majority density normalized by the
reference density, $n_0$.\label{f:profiles}}
\end{center}
\end{figure}

We estimate the zero-temperature equation of state by analyzing the
density profiles obtained at the lowest temperature ($T/T_F \approx 0.03$
where $T_F$ is the Fermi temperature of the majority component) in
Ref.~\cite{SSS08}. A spin mixture of the two lowest hyperfine states of
$^6$Li atoms was prepared in a 3D harmonic trap on a broad Feshbach
resonance, located at a magnetic field of 834~G~\cite{BAR04a}, resonantly
enhancing the interactions between the two spin states. The detailed
description of the experimental procedure for the sample preparation and
the signal processing has been provided in Ref.~\cite{SSS08}.

Figure~\ref{f:profiles} shows the spatial structure of a resonantly
interacting Fermi mixture in a harmonic trap. According to the
zero-temperature phase diagram~\cite{SSS08}, three distinctive spatial
regions can be identified in the inhomogeneous sample with population
imbalance ($y_0 < 1$). When the chemical potential ratio at center $y_0$
is larger than a critical value $y_c$ for the superfluid-to-normal phase
transition, the sample has (I) a fully-paired superfluid core ($x=1$,
$y>y_c$) surrounded by (II) a partially-polarized normal gas ($0< x\leq
x_c$, $y<y_c$), showing a discontinuity in the density ratio $x$ at the
phase boundary. In the outer region where the minority component is
completely depleted, (III) a fully-polarized non-interacting Fermi gas
forms ($x=0$, $y<y_m<y_c$). The radii, $R_c$ and $R_2$ are the I-II and
II-III transition points, defining $y(R_c)=y_c$ and $y(R_2)=y_m$,
respectively.

The non-interacting ideal Fermi gas in the outer $(r>R_2)$ region
provides a reliable method to measure the local chemical potential
$\mu_1(r)$ in the strongly interacting, inner $(r<R_2)$ region. Since
$\varepsilon_{F1}=\mu_1$ in the outer region, the extension of the
non-interacting ideal Fermi distribution $n_0(r)$, fit to the outer
region, into the inner region gives the local chemical potential as
$\mu_1(r)= \alpha n_0^{2/3}$, consequently $\mu_1/\varepsilon_{F1}=
(n_0/n_1)^{2/3}$. This method allows to measure the equation of state
directly from the shape of the cloud without any absolute calibration for
particle density. Furthermore, when a sample has a superfluid core, i.e.
the whole range of the density ratio $0 \leq x \leq 1$, a single shot
image of the sample can provide all information for the determination of
the equation of state.

In the experiments, the total population imbalance was controlled to be
$\delta=(N_1-N_2)/(N_1+N_2)=44(4)\%$ less than the critical imbalance
$\delta_c(\approx 75\%)$~\cite{ZSS06a,SZS06} to have a superfluid core,
i.e. $y_0>y_c$, where $N_i$ is the total atom number of component $i$.
The phase boundary $R_c$, located by the kink and peak position in the
column density difference profile, was measured to be $R_c/R_1=0.430(3)$
and the critical density ratio was measured to be $x_c=0.53(5)$. The
reference density $n_0(r)$ and the radius $R_1$$(R_2)$ were determined
from the fit of the outer region, $r>R_2$$(r>R_c)$ of the
majority(minority) column density profile to a zero-temperature
Thomas-Fermi (TF) distribution.

\begin{figure}
\begin{center}
\includegraphics[width=3.35in]{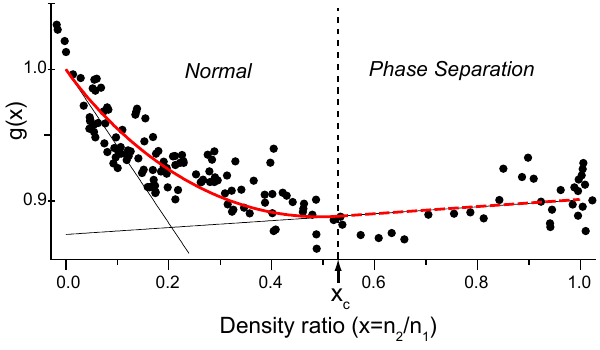}
\caption{(Color online) Thermodynamic potential at unitarity. The
universal function $g(x)$ for energy density was constructed as a
function of the density ratio $x$ of the two components, from five
independent measurements of the density profiles (Fig.~\ref{f:profiles}).
The critical chemical potential ratio and the critical density ratio were
measured to be $y_c=0.03(2)$ and $x_c=0.53(5)$, respectively (see text
for determination method). The red (dark gray) solid line is obtained by
fitting the model equation Eq.~(\ref{e:model}) with $\{\xi, y_c,
y_m,x_c\}=\{0.42, 0.03, -0.58, 0.53\}$ to the normal region ($x<x_c$) of
$g(x)$. A red (dark gray) dashed line in the phase separation region
($x_c<x<1$) connects the two points ($x=1$, $g(1)=(2\xi)^{3/5}$) and
($x=x_c$, $g(x_c)$).\label{f:function}}
\end{center}
\end{figure}

The well-known zero-temperature thermodynamics of a balanced superfluid
and a non-interacting Fermi gas provide physical constraints on the form
of the universal function. Since the chemical potential of a fully-paired
($x=1$) superfluid is proportional to the Fermi energy, i.e.
$\mu_s=(\mu_1+\mu_2)/2=\xi \varepsilon_{F1}$, we have
$\mu_1/\varepsilon_{F1}= 2\xi/(1+y)$ and $g(1)=(2\xi)^{3/5}$. On the
other hand, a fully-polarized ($x=0$) non-interacting Fermi gas has
$\mu_1=\varepsilon_{F1}$ so that $g(0)=1$. The universal parameter $\xi$
can be independently determined from the majority profile by comparing the
curvature of the Fermi energy distribution $\varepsilon_{F1}(r)\propto
n_1(r)^{2/3}$ in the core region and in the outer region, as $\xi = (d^2
\varepsilon_{F1} / d r^2)_{r>R_2} / (d^2 \varepsilon_{F1} / d
r^2)_{r<R_c}$. This determination is, however, technically limited due to
the low signal-to-noise ratio. In the following analysis we use the
theoretically predicted value $\xi_{th}=0.42(1)$~\cite{ABC04,CR05},
confirmed in previous
measurements~\cite{OHG02,GHG03,BAR04b,BKC04,KTT05,PLK06,SGR06,SSS07b}.

Figure~\ref{f:function} displays the universal function $g(x)$
constructed from the density profiles at the lowest temperature. The
critical value $y_c$ was determined to be $y_c=0.03(2)$ such that the
average value of $g(x)$ for $x>0.9$ gives $(2\xi_{th})^{3/5}=0.90$. This
critical value $y_c$ has been discussed in Ref.~\cite{SSS08} to
demonstrate the stability of a fully-paired superfluid state at zero
temperature. At zero temperature, $g(x)$ is not defined for a homogeneous
system with $x_c<x<1$. The sparse population of the data points in the
region of $x_c<x<1$ indicates the phase separation in the sample,
associated with the first-order phase transition.

\begin{figure}
\begin{center}
\includegraphics[width=3.3in]{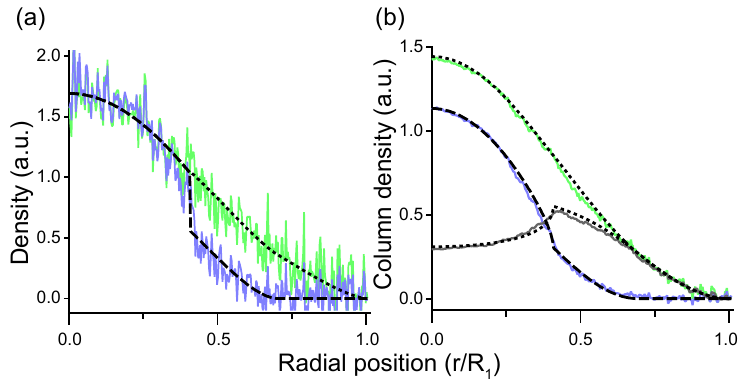}
\caption{(Color online) Comparison between the experimental data and the
model. (a) The 3D density profiles and (b) the column density profiles for
$\delta=44\%$ (dot-dashed: majority, dashed: minority, dotted:
difference) generated from the model equation of state (the red (dark
gray) line in Fig.~\ref{f:function}). Experimental data in
Fig.~\ref{f:profiles} are plotted together for comparison (same color
usage).\label{f:comp}}
\end{center}
\end{figure}

The other critical value $y_m$ represents the binding energy $E_b$ of a
single minority resonantly interacting with a majority Fermi sea as $E_b
= \lim_{x\rightarrow 0^+} \mu_2=y_m \varepsilon_{F1}$.
$y=g'(x)/(g(x)-xg'(x))$ and $g(0)=1$ give $y_m=g'(0)$. By fitting $1+y_m
x$ to $g(x)$ for $x<0.1$, we estimated $y_m=-0.58(5)$, which is in good
agreement with the recent theoretical results of $-0.6$~\cite{Chevy06},
$-0.58(1)$~\cite{LRG06}, $-0.6066$~\cite{CRL07}, $-0.54(4)$~\cite{BF07},
and $-0.618$~\cite{PS07}. From the definitions, $y_c=y(R_c)$ and
$y_m=y(R_2)$, the estimated critical values $\{y_c, y_m\}=\{0.03(2),
-0.58(5)\}$ suggest $R_2=0.707(20) R_1$ for $R_c=0.43 R_1$. However, the
minority radius was measured to be $R_2/R_1=0.73(1)$, suggesting
$y_m=-0.69(8)$. We attribute this discrepancy to the fact that the shape
of the minority density profile in the intermediate region cannot be
completely captured with a zero-temperature TF distribution~\cite{note1}.

Following a normal Fermi liquid description~\cite{LRG06, CRL07, PS07}, we
consider a model for the equation of state of a partially polarized Fermi
gas in the following form,
\begin{equation}
g(x)^{5/3}=1+\frac{5}{3} y_m x + \frac{1+c x}{m^*} x^\gamma~~\mathrm{for}~
x<x_c.\label{e:model}
\end{equation}
satisfying the boundary conditions $g(0)=1$ and $g'(0)=y_m$ at $x=0$. The
second term corresponds to the momentum-independent binding energy for
the minority atoms and the third term describes the deviation from the
free particle behavior, regarding $m^*$ as the effective mass and $c$ as
its correction (for the non-interacting case, $g_0(x)^{5/3}=1+x^{5/3}$).
The equilibrium condition for the coexistence of two spatially separate
phases requires that the two phases have the same chemical potential and
pressure at the critical point~\cite{note2}, imposing the boundary
conditions at $x=x_c$ such as $g(x_c)=[(1+x_c y_c)/(1+y_c)]g(1)$ and
$g'(x_c)=[y_c/(1+y_c)]g(1)$. Then, for given values
$\{\xi,y_c,y_m,x_c\}$, this model relies on only one free parameter. With
$\{\xi, y_c, y_m,x_c\}=\{0.42, 0.03, -0.58, 0.53\}$, the fit of
Eq.~(\ref{e:model}) to the intermediate region ($x<x_c$) gives
$\gamma=1.60(13)$, having $m^*=1.06$ and $c=-0.019$. Quantum Monte-Carlo
calculations for small $x$ predicts $m^*=1.04(3)$ with
$\gamma=5/3$~\cite{LRG06}, which is very close to the observed behavior.
Figure~\ref{f:comp} displays the density profiles generated from the
model equation of state, together with the experimental data.

Our observation of $\gamma \approx 5/3$ and the small change in the
effective mass suggests that a polarized Fermi gas with resonant
interactions can be described as a normal Fermi liquid with weakly
interacting quasiparticles. However, it is an open question whether the
Fermi liquid description is still valid for high minority concentrations,
where the Pauli blocking effect of the minority Fermi sea might play an
important role. The possibility of the exotic ground state of a partially
polarized system has been suggested by the recent observation of the
temperature-dependent spectral shift in the minority rf excitation
spectrum~\cite{SSS07}. We note that it is not clear how to distinguish
possible exotic states~\cite{BFS06} via the equation of state. More
experimental studies for microscopic properties of the system, e.g.
majority rf spectroscopy, are necessary to clarify the issue.

In conclusion, we measure the equation of state of a two-component Fermi
gas with resonant interactions by analyzing the \textit{in situ} density
distributions of the trapped sample. In a similar way, the density
profiles at finite temperature may reveal the excitation spectrum of the
system~\cite{CR07}.

The author thanks C. H. Schunck and A. Schirotzek for experimental
assistance, and W. Ketterle, A. Bulgac, M. M. Forbes, C. Lobo, and S.
Reddy for discussions. This work was supported by NSF, ONR, MURI, and
DARPA.


\begin{thebibliography}{29}
\bibitem{note1}
The critical values $y_c$ and $y_m$ are related with the radii $R_1$,
$R_2$, and $R_c$ as
$\kappa\equiv(1-y_c)/(1-y_m)=(R_1^2-R_2^2)/(R_1^2-R_c^2)$. Our measured
critical values $\{y_c, y_m\}=\{0.03, -0.58\}$ suggest $\kappa \approx
0.61$. The experimental data of the radii reported in Ref.~\cite{ZSS06b}
show $\kappa\approx 0.70$, where the expansion dynamics was involved.
\bibitem{note2}
The pressures of a fully-paired superfluid is given as
$\mathcal{P}_\mathrm{S}=\frac{4}{5} \alpha^{-3/2} \xi^{-3/2}
\mu_s^{5/2}$. The pressure of a partially-polarized normal gas can be
expressed as $\mathcal{P}_\mathrm{N}=\frac{2}{5} \alpha^{-3/2}[\mu_1/(g-x
g')]^{5/2}$~\cite{BF07}. $\mathcal{P}_\mathrm{S}=\mathcal{P}_\mathrm{N}$
at $x=x_c$ gives $g'(x_c)=(g(1)-g(x_c))/(1-x_c)$ using
$y_c=g'(x_c)/(g(x_c)-x_cg'(x_c))$ and $\mu_s=(\mu_1+\mu_2)/2$.
\end{thebibliography}

\end{document}